\documentclass[
aps,
prl,
twocolumn,
preprintnumbers,
amsmath,
amssymb,
superscriptaddress,
nofootinbib,
nobibnotes
]{revtex4-2}

\usepackage{graphicx}
\usepackage{bm}
\usepackage{hyperref}
\usepackage{xcolor}
\usepackage{IEEEtrantools}

\usepackage{microtype}

\usepackage{float}

\restylefloat{table}

\begin{document}

\title{Magnetic-field-tunable cyclotron hyperbolic polaritons}

\author{Zijian Zhou}
\affiliation{Department of Physics and Astronomy, Stony Brook University; Stony Brook, New York 11794, USA}

\author{Ran Jing}
\affiliation{Department of Physics and Astronomy, Stony Brook University; Stony Brook, New York 11794, USA}

\author{Heng Wang}
\affiliation{Department of Physics and Astronomy, Stony Brook University; Stony Brook, New York 11794, USA}

\author{Lukas Wehmeier}
\affiliation{National Synchrotron Light Source II, Brookhaven National Laboratory, Upton, New York 11973, USA}

\author{Mengkun Liu}
\affiliation{Department of Physics and Astronomy, Stony Brook University; Stony Brook, New York 11794, USA}

\author{Bing Cheng}
\thanks{Contact author: chengbing986@gmail.com}
\affiliation{Department of Physics and Astronomy, Stony Brook University; Stony Brook, New York 11794, USA}
\affiliation{Research Laboratory of Electronics, Massachusetts Institute of Technology, Cambridge, MA 02139, USA}
\date{\today}

\date{\today}

\begin{abstract}

Hyperbolic polaritons are conventionally associated with structural anisotropy or phononic Reststrahlen bands. Here, we predict a new class of hyperbolic polaritons arising from magnetic-field-induced cyclotron motion of charge carriers. When a perpendicular magnetic field is applied to high-mobility semimetals, the cyclotron response drives the in-plane dielectric function from metallic- to insulating-like below the cyclotron resonance frequency, while the out-of-plane response remains metallic. This anisotropy creates a hyperbolic dielectric environment that supports field-tunable hyperbolic polaritons. We develop a comprehensive theoretical framework incorporating coupling to other collective excitations and show that these modes can be directly visualized in real space via terahertz near-field nanoscopy. Our work identifies cyclotron motion as a new route to hyperbolic polaritons and establishes a versatile platform for magnetically programmable nanophotonics.

\end{abstract}

\maketitle

\setlength{\parskip}{0.1em}

In anisotropic media, when the dielectric anisotropy becomes sufficiently strong such that different principal components of the dielectric tensor acquire opposite signs, the dispersion relation of electromagnetic waves in $k$-space will develop open hyperboloid isofrequency surfaces \cite{HP_Basov_review,HP_Koppens_review}. As a consequence, hybrid light–matter excitations known as hyperbolic polaritons can emerge and propagate within the medium. These modes exhibit several distinctive properties: their hyperbolic dispersion supports unbounded in-plane momenta, a strongly enhanced photonic density of states, and highly directional energy flow, enabling ray-like propagation and extreme electromagnetic confinement beyond what is achievable in conventional polaritonic systems \cite{Caldwell2014,Caldwell2019_hBN_review}. Owing to these attributes, hyperbolic polaritons have recently emerged as a central platform in nanophotonics, particularly in cavity quantum electrodynamics (cavity-QED), where deep-subwavelength confinement and large mode densities can enable strong and even ultrastrong light–matter coupling \cite{Roadmap2DPhotonics2025,Basov_review_2025}. In such regimes, enhanced electromagnetic fluctuations and virtual-photon exchange can mediate non-perturbative interactions, offering pathways to renormalizing electronic excitation spectra and modifying many-body ground-state properties even in the absence of external optical pumping \cite{CavityQED_2023,Cavity_2025,Kipp2025CavityVdW,Keren2025CavityAlteredSC}.

Experimentally, van der Waals crystals such as hexagonal boron nitride (hBN) \cite{hBN_HP_2014}, $\alpha$-molybdenum trioxide ($\alpha$-MoO$_3$) \cite{Ma2018_MoO3}, and their derivatives have served as prototypical platforms for realizing and manipulating hyperbolic polaritons \cite{Dai2015_hBN_gra,Woessner2015_hBN_gra,Hu2022_Gra_MoO3,Ruta2022,Sternbach2023,Lukas2024}. In these systems, the hyperbolic response originates primarily from strong phonon anisotropy, giving rise to hyperbolic phonon polaritons. More recently, attention has expanded to materials in which hyperbolicity originates from anisotropic electronic structures rather than phononic responses. Representative examples include the nodal-line semimetal ZrSiSe and the correlated van der Waals metal MoOCl$_2$, which host hyperbolic plasmon polaritons \cite{HP_nodal_metal_2022,HP_metal_2025,Venturi2024}. Despite these advances, most hyperbolic polaritons reported to date remain fundamentally tied to intrinsic lattice or electronic anisotropies encoded in the crystal and band structures, confining hyperbolicity to fixed spectral windows. As a result, such hyperbolic modes are difficult to actively tune using external control parameters, such as electric or magnetic fields. Overcoming this limitation represents both a central challenge and a key opportunity for advancing tunable hyperbolic photonics and cavity-QED–enabled platforms.

\setlength{\belowcaptionskip}{-0.5cm} 

\begin{figure*}[t]
\includegraphics[width=0.96\textwidth]{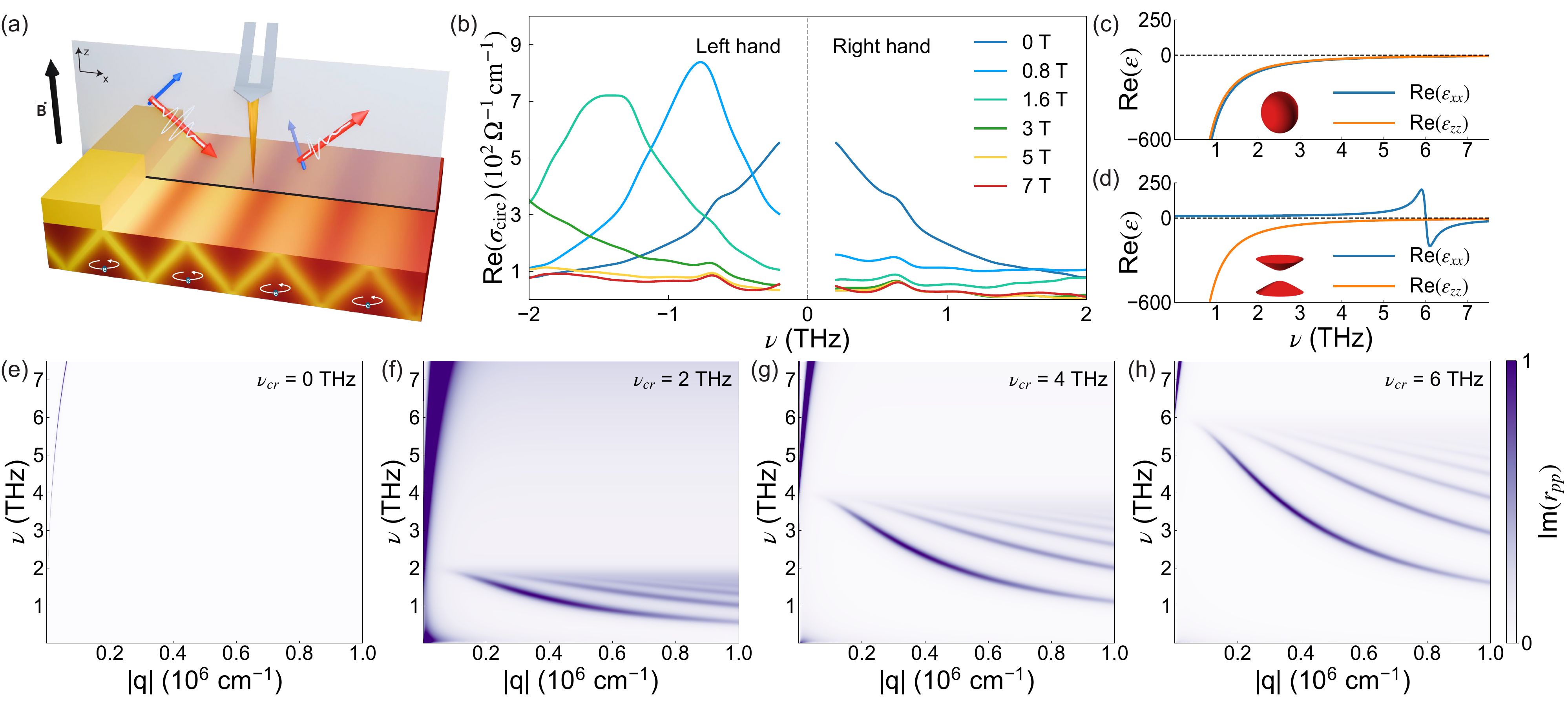}
\caption{\label{fig:Bi50nm_THz_pump_long} Magnetic-field-driven cyclotron hyperbolic polaritons. (a) Schematic of magnetic-field–driven terahertz hyperbolic polaritons in a high-mobility semimetal. A magnetic field applied along the $z$-axis confines electron cyclotron motion to the $xy$-plane. A deposited gold film launches the polariton mode, which can be detected by terahertz scattering-type near-field microscopy through a line scan along the black line. (b) Real part of the circularly polarized terahertz conductivity $\sigma_{\mathrm{circ}}$ of Cd$_3$As$_2$ thin films under magnetic field. (c,d) Simulated Re($\epsilon_{xx}$) and Re($\epsilon_{zz}$) at zero and finite magnetic fields up to 7.5 THz. Insets show the corresponding isofrequency contours at 2.5 THz. (e–h) Numerical simulations of reflection coefficient Im($r_{pp}$) for a 100-nm-thick semimetal film with cyclotron frequencies $v_{\mathrm{cr}}$ = 0, 2, 4, and 6 THz. The dark branches (poles in $r_{pp}$) reveal the hyperbolic polariton modes. The color scale has been normalized by the maximum intensity of the fundermantal branch. }
\label{Fig1}
\end{figure*}

In this work, we demonstrate a new class of hyperbolic polaritons that can be realized and tuned by magnetic fields in high-mobility semimetals. We develop a full theoretical treatment for this phenomenon and elucidate how the Hall response and lattice vibrations influence the polariton dispersion. To establish the basic physics in a general framework, we consider a high-mobility semimetal thin film containing light, weakly scattered carriers. In the low-energy regime (below 7 THz), its optical response, such as the optical conductivity $\sigma(\omega)$, is described by the Drude model: $\sigma(\omega)$=$\epsilon_0$$\omega_p^2\tau/(1-i\omega\tau)$, where $1/2\pi\tau$ and $\omega_p$ are Drude scattering rate and plasma frequency, respectively. Here we temporarily neglect the possible phonon response and will discuss its influence at a later stage. At zero magnetic field, the low-energy optical response along all crystal axes (labelled as $x$, $y$, and $z$) is Drude-like. As a result, the real part of the dielectric function, $\epsilon(\omega)$ $=$ 1 + $i\sigma(\omega)$$/\epsilon_{0}\omega$, is negative in all directions, which therefore cannot support hyperbolic polaritons. 

The presence of magnetic fields, however, can completely change this landscape. When a magnetic field is applied perpendicular to the thin film shown in Fig. \ref{Fig1}a, the electrons undergo cyclotron motion in the $xy$ plane. This motion gives rise to a cyclotron resonance (CR) mode in the optical response \cite{Bi2Se3_CR,CdAs_IR_magneto2016,PbSnTe_CR,CdAs_CR}. Here we show an experimental example. Figure \ref{Fig1}b displays the real part of circularly polarized terahertz conductivity Re($\sigma_{\mathrm{circ}}$) of Cd$_3$As$_2$ thin films measured by time-domain magneto-terahertz spectroscopy.  The measurement details are provided in the Supplementary Material. At zero field, Re($\sigma_{\mathrm{circ}}$) exhibits a Drude-like peak at zero frequency. As the magnetic field increases, the peak shifts to finite negative frequency (left-hand) and the positive-frequency (right-hand) response diminishes, a behavior characteristic of CR mode. Once the field exceeds 3 T, the CR mode shifts to frequencies well above 2 THz. In principle, this CR mode can be viewed as an in-plane “electronic phonon”, analogous to the phonon resonance in hBN but tunable by the magnetic field. When this mode is sufficiently sharp, it drives the real part of in-plane dielectric function Re($\epsilon_{xx}$) to change sign from negative (metallic-like) to positive (insulating-like) below the CR frequency. Re($\epsilon_{xx}$) of Cd$_3$As$_2$ thin films is provided in the Supplementary Material. In contrast, the optical response along the magnetic-field direction remains nearly unchanged, as carriers moving parallel to the field experience no Lorentz force, and may even become more metallic via mechanisms like the chiral magnetic effect \cite{Ong2015,Li02015,Cheng_2021}. Consequently, below the CR frequency, the in-plane and out-of-plane dielectric functions acquire opposite signs—specifically, Re($\epsilon_{xx}$)$>$0 and Re($\epsilon_{zz}$)$<$0—a condition that supports type-I hyperbolic polaritons. To illustrate this behavior within a general framework applicable to high-mobility semimetals and not limited to Cd$_3$As$_2$, Fig. \ref{Fig1}c and \ref{Fig1}d present the simulated Re($\epsilon_{xx}$) and Re($\epsilon_{zz}$) over a frequency range up to 7.5 THz, under zero and finite magnetic field, respectively (see Supplementary Material for details). In the following, we use the CR frequency—rather than the magnetic field strength—to characterize the field-induced cyclotron motion of electrons, since its field dependence varies among semimetals. For a representative case with CR frequency $v_{\mathrm{cr}}$ = 6 THz, the isofrequency contour of light propagating in the film becomes hyperbolic below $v_{\mathrm{cr}}$—an effect entirely absent in the zero-field case.

Despite this exciting scenario, the cyclotron motion of carriers also introduces a Hall component, represented by the off-diagonal term $\epsilon_{xy}$, into the full dielectric tensor. This raises the question of whether $\epsilon_{xy}$ might smear out or otherwise modify the conditions required for sustaining hyperbolic polaritons. Hence, to solve the field-driven hyperbolic polariton problem, one must fully account for the 3 $\times$ 3 dielectric tensor: 

 \vspace{-0.3cm}

\begin{equation}
\epsilon = \begin{pmatrix}
\epsilon_{xx} & \epsilon_{xy} & 0 \\
-\epsilon_{xy} & \epsilon_{yy} & 0 \\
0 & 0 & \epsilon_{zz}
\end{pmatrix}
\end{equation}

\noindent To simplify the discussion, we assume the diagonal in-plane dielectric response is isotropic, i.e. $\epsilon_{xx}=\epsilon_{yy}$. Without loss of generality, we take $xz$ plane as the plane of incidence and start from the general solution of the electromagnetic wave propagating in the film \cite{GoncalvesPeres2016}: $E$ = $(E_x, E_y,E_z)$ $e^{i(qx+kz)}e^{-i\omega t}$; $B$ = $(B_x, B_y,B_z)$ $e^{i(qx+kz)}e^{-i\omega t}$. Here $q$ and $k$ are the $x$- and $z$-components of the wave vector, respectively. We incorporate the dielectric tensor into Maxwell’s equations and find that $E_x$ and $E_y$ of eigenmodes inside the film are coupled: 

 \vspace{-0.3cm}

\begin{equation}
[k_0^2\epsilon_{xx}-(q^2+k^2)]E_y=k_0^2\epsilon_{xy}E_x
\label{TI}
\end{equation}

\setlength{\belowcaptionskip}{-0.5cm} 

\begin{figure}[t]
\includegraphics[width=0.47\textwidth]{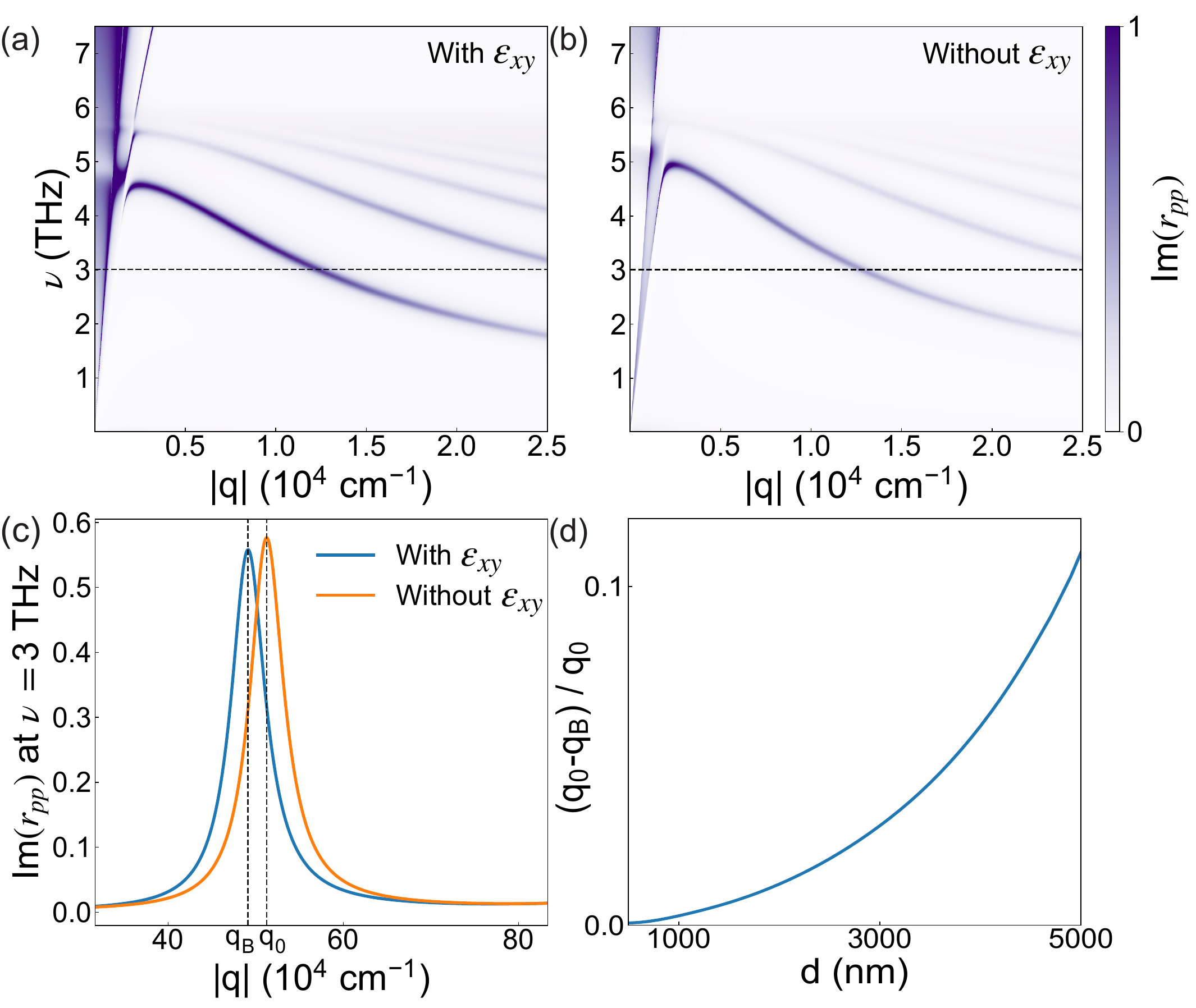}
\caption{\label{fig:Bi50nm_THz_pump_long}  Influence of Hall response on the polariton dispersion. (a,b) Numerical simulations of the reflection coefficient $r_{pp}$ for a 3500-nm-thick semimetal film computed at $v_{\mathrm{cr}}$ = 6 THz, with and without the Hall component $\epsilon_{xy}$. (c) Representative line cut at $v$ = 3 THz, taken along the dashed black lines in panels (a) and (b). (d) Thickness dependence of the shift ratio induced by the Hall response $\epsilon_{xy}$.}
\label{Fig2}
\end{figure}

\noindent Here $k_0=\omega/c$ is the wave vector of light in vacuum. This equation means that the field-driven eigenmodes cannot be simply classified as $p$- and $s$-polarized waves as in conventional hyperbolic crystals, e.g., hBN and $\alpha$-MoO$_3$. The effect of $\epsilon_{xy}$—or equivalently, a magnetic field—can be viewed as effectively rotating the in-plane polarization of the eigenmodes by a complex angle $\arctan(E_y/E_x)$. The $k$ vectors of two eigenmodes are given by:

\vspace{-0.3cm}

\begin{equation}
k_{e/o}^2 = \tfrac{1}{2}\!\left[(k_{e0}^2 + k_{o0}^2)
\pm \sqrt{(k_{e0}^2 - k_{o0}^2)^2 - 4a}\,\right]
\label{TI}
\end{equation}

 \noindent Here, for conciseness, we continue to denote $k_e$ and $k_o$ as the wave vectors of the extraordinary- and ordinary-like modes. The parameters $k_{e0}$, $k_{o0}$, and $a$ are defined as $k_{e0}$ = $\sqrt{k_0^2\epsilon_{xx}-q^2\epsilon_{xx}/\epsilon_{zz}}$, $k_{o0}$ = $\sqrt{k_0^2\epsilon_{xx}-q^2}$, and $a=\epsilon_{xy}^2k_0^2(k_0^2-q^2/\epsilon_{zz}^2)$. If $\epsilon_{xy}=0$, the equations mathematically reduce to those of the standard uniaxial phonon-polariton case, as in hBN \cite{hBN_HP_2014}. Details of the derivation are provided in the Supplementary Material.

With above equations and the boundary conditions, we can directly calculate the reflection  coefficient of the films using transfer matrix method. The poles of the reflection  coefficient reveal the polariton modes. For simplicity, we consider a $p$-polarized plane wave, $E$ = $(E_x,0,E_z)$ $e^{i(qx+kz)}e^{-i\omega t}$, incident on the thin film as shown in Fig. \ref{Fig1}a.  In the presence of magnetic field, Eq. (2) indicates that an incident $p$-polarized wave inevitably generates an $s$-polarized component through the cyclotron motion of charge carriers. As a result, the reflection  coefficient has two components, $r_{pp}$ and $r_{ps}$, representing reflections into $p$- and $s$-polarized light, respectively. The poles of these two coefficients yield the same dispersion for the field-driven hyperbolic polaritons. The details of calculation are provided in the Supplementary Material.

\begin{figure}[t]
\centering
\includegraphics[width=0.49\textwidth]{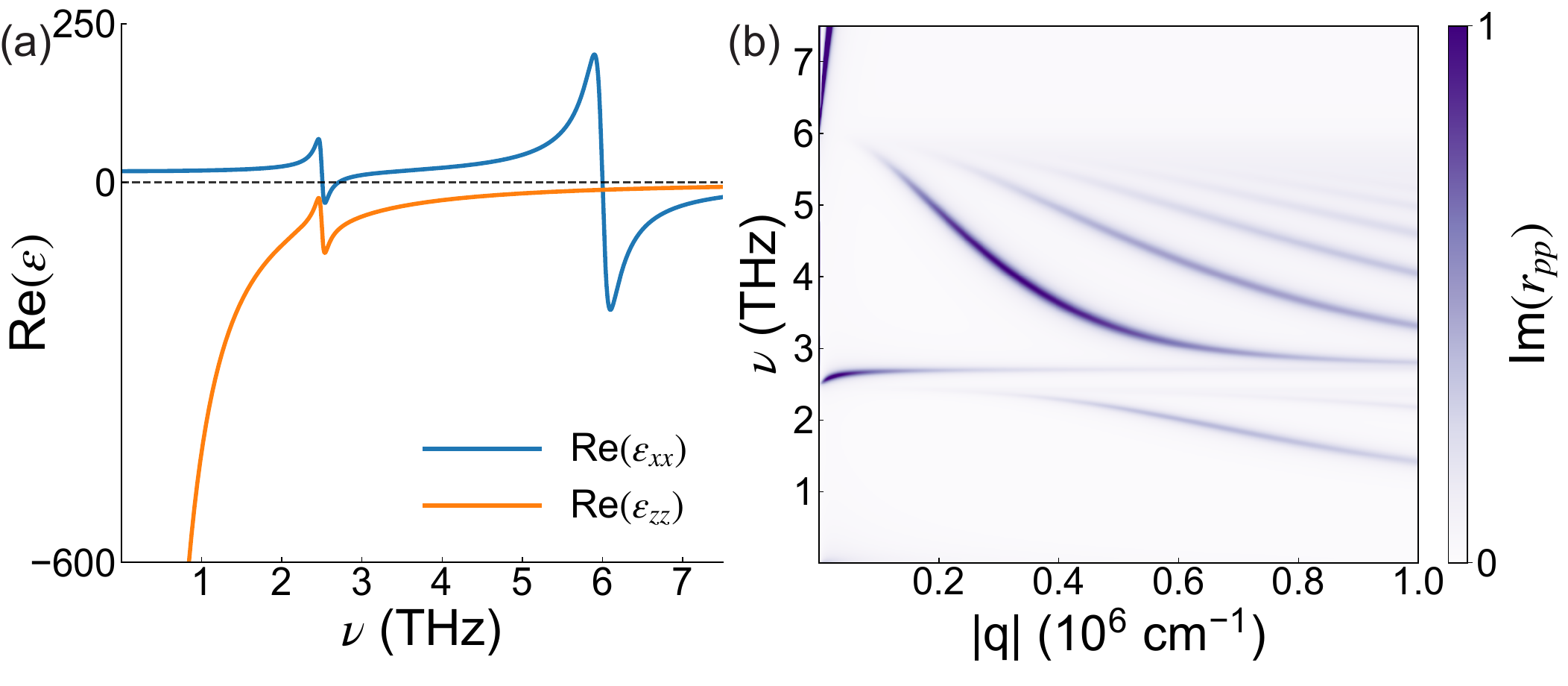}
\caption{\label{fig:Bi50nm_THz_pump_long} Coupling between field-driven cyclotron hyperbolic polaritons and lattice vibrations. (a) Numerical simulations of the dielectric function including both an IR-active phonon at 2.5 THz and a cyclotron resonance mode at 6 THz. (b) The numerical simulations of polariton dispersion for a 100-nm-thick semimetal film including the phonon effect.}
\label{Fig3}
\end{figure} 

Figure \ref{Fig1}e–\ref{Fig1}h show the numerical simulations of Im($r_{pp}$) for a 100-nm-thick semimetal film at CR frequencies $v_{\mathrm{cr}}$ = 0, 2, 4, and 6 THz. Simulation parameters and details are provided in the Supplementary Material. At zero CR frequency, no hyperbolic modes are observed, as expected, apart from the ordinary surface plasmon polaritons that appear in the low-momentum regime, which is not the focus of this work. With increasing $v_{\mathrm{cr}}$, multiple branches of hyperbolic polaritons emerge and become pronounced. Their dispersion shows a decrease in energy with increasing in-plane momentum $\lvert q \rvert$, confirming that these modes correspond to type-I hyperbolic polaritons. As $v_{\mathrm{cr}}$ increases further, these modes extend toward higher frequencies but terminate at $v_{\mathrm{cr}}$, beyond which they vanish as the field-induced electronic hyperbolicity disappears. The CR frequency therefore defines the upper frequency bound of the hyperbolic polaritons and scales directly with magnetic field, establishing it as the key control parameter governing the existence and tunability of hyperbolic polaritons in semimetal films.

\begin{figure*}[t]
\centering
\includegraphics[width=0.99\textwidth]{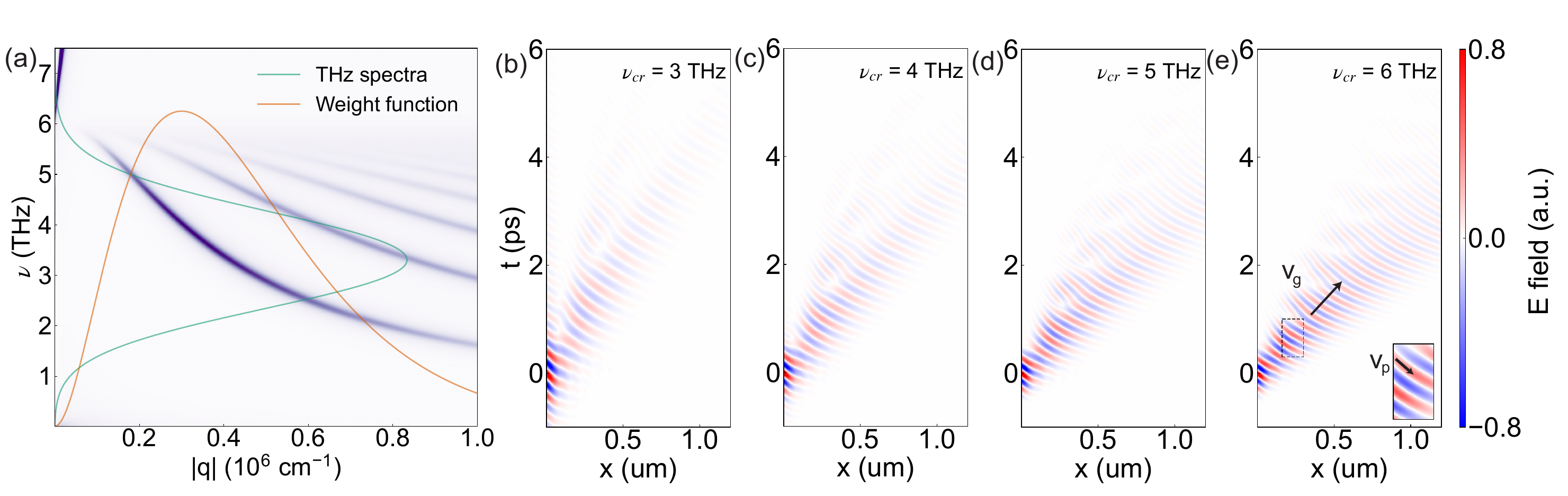}
\caption{\label{fig:Bi50nm_THz_pump_long} Spacetime map of the field-driven cyclotron hyperbolic polaritons. (a) The THz pulse's spectrum, and momentum-coupling weight function which describes the near-field probe’s coupling strength to momentum used in the simulation of spacetime map. (b-d) The simulated spacetime maps of the hyperbolic polariton wave packet for CR frequency $v_{\mathrm{cr}}$ = 3, 4, 5, and 6 THz. The inset highlights the fixed-phase wavefronts. The black line with arrows indicates the directions of the group velocity $v_{\mathrm{g}}$ and phase velocity  $v_{\mathrm{p}}$, respectively.}
\label{Fig4}
\end{figure*}

After demonstrating field-driven hyperbolic polaritons in semimetals, a key question arises: does the Hall response $\epsilon_{xy}$ indeed participate in their formation and dispersion? To address this, we performed simulations at $v_{\mathrm{cr}}$ = 6 THz, explicitly comparing results obtained with and without the Hall dielectric term $\epsilon_{xy}$. Our results show that when the film is sufficiently thin, the Hall term has little influence on the dispersion. However, as the film becomes thicker, the Hall contribution grows and begins to noticeably alter the hyperbolic polariton dispersion. The corresponding results for a thick film (3500 nm) are shown in Fig. \ref{Fig2}a and \ref{Fig2}b. Both simulations exhibit well-defined hyperbolic polariton modes; however, the inclusion of $\epsilon_{xy}$ introduces clear modifications in the low-$\lvert q \rvert$ regime. In this region, the modes show a pronounced downshift in frequency at a given $\lvert q \rvert$ compared to the case without $\epsilon_{xy}$. This comparison demonstrates that the Hall response indeed modifies the dispersion of hyperbolic polaritons. To make this difference more evident, a representative line cut at $v$ = 3 THz is shown in Fig. \ref{Fig2}c, where the fundamental branch reveals a pronounced shift towards lower momentum. The dependence of this behavior on film thickness is summarized in Fig. \ref{Fig2}d, which is consistent with the trend described above.

Having established the role of Hall response, we next examine another key factor that can influence the polariton behavior—the coupling between hyperbolic modes and lattice vibrations. In semimetals, infrared-active phonons are commonly found in the terahertz range \cite{WTe2_IR_phonon_2015,CdAs_IR_magneto2016,ZrTe5_phonon_2018}. For simplicity and without loss of generality, we consider an IR-active phonon at 2.5 THz and simulate the dielectric function including both the phonon and a CR mode at 6 THz. The results in Fig. \ref{Fig3}a show that, in addition to the feature that the CR mode drives Re($\epsilon_{xx}$) from negative to positive below $v_{\mathrm{cr}}$, the phonon further reshapes Re($\epsilon_{xx}$) and drives it from positive to negative near the phonon frequency. Figure \ref{Fig3}b shows the simulated polariton dispersion including the phonon effect. The phonon markedly modifies the dispersion: the original hyperbolic polariton branches terminate near the phonon frequency, and a new branch emerges below it. The results demonstrate that coupling to lattice vibrations provides an additional route to tailor the dispersion of field-driven hyperbolic polaritons.

To experimentally detect the proposed field-driven hyperbolic polaritons, terahertz scattering-type scanning near-field optical microscopy (THz-sSNOM) provides a powerful approach \cite{Map_SPP_THz_2024}. This technique enables direct visualization of polaritonic wave packets in the time domain and resolves their phase and group velocities. To simulate the features observable by THz-sSNOM, we consider a geometry in which the hyperbolic polaritons are launched by a gold film deposited on the edge of a half-plane semimetal thin film, as illustrated in Fig. \ref{Fig1}(a). When a magnetic field is applied, the THz pulse launches hyperbolic polaritons at the gold edge, which then propagate perpendicular to the edge. The tip-scattered near-field signal is recorded as the tip performs a line scan (the black line in Fig. \ref{Fig1}(a)) perpendicular to the gold edge while the time delay $t$ is varied, yielding a spacetime map that displays a characteristic fringe pattern. Figure \ref{Fig4}a shows the incident THz spectrum and the momentum-coupling weight function used in the simulations, peaking at 3 THz and $0.3\times10^6$ cm$^{-1}$. Tailoring either enables selective access to different polaritonic energies and momenta, e.g., by tuning the tip-apex radius $r$ that controls the near-field tip-momentum coupling centered at $q \sim 1/r$, without qualitatively altering the main simulation features. Figure \ref{Fig4}b–\ref{Fig4}d present the simulated spacetime maps of the hyperbolic polariton wave packet in a 100-nm-thick semimetal film for CR frequencies $v_{\mathrm{cr}}$ = 3, 4, 5, and 6 THz. The maps clearly show polariton wave packets propagating along the $+x$ direction with a positive slope in the $tx$ plane. As these wave packets correspond to the flow of energy launched from the sample edge, their propagation defines a positive group velocity, following the convention of previous infrared studies \cite{Map_HP_2015}. A key observation is that the slope of the trajectory in the spacetime maps, $dx/dt$, gradually increases with magnetic-field strength, indicating that the group velocity $v_g$ grows as the field increases. This trend is fully consistent with the polariton dispersion in Fig. \ref{Fig1}, where the group velocity determined by $\lvert dv/dq \rvert$ also increases with the magnetic field. Interestingly, as shown in the inset of Fig. \ref{Fig4}d, the fixed-phase wavefronts propagate with a negative slope in the $tx$ plane, indicating a negative phase velocity in the spacetime map. The opposite directions of phase and energy flow, directly accessible via THz-sSNOM, provide clear evidence of a type-I polariton mode even without measuring its full dispersion.

The magnetic-field–driven cyclotron hyperbolic polaritons proposed here substantially expand the landscape of polariton physics by introducing a dynamically programmable, electronic route to hyperbolicity. Our mechanism is general and broadly applicable to high-mobility semimetals or semiconductors in thin-film geometries, requiring only the presence of a sharp cyclotron resonance in the terahertz frequency range. Owing to recent advances in topological semimetal research, a wide class of experimentally accessible systems already meets these criteria \cite{Cd3As22013,Liu2014,Suyang2015,topo_semimetal_review}. Importantly, the required magnetic fields are moderate and compatible with standard laboratory environments, while the resulting hyperbolic modes are continuously tunable over a broad terahertz bandwidth, in stark contrast to the lattice-defined Reststrahlen bands of conventional hyperbolic crystals.

Beyond tunability, cyclotron-driven hyperbolic polaritons provide a direct and sensitive probe of electronic structure\cite{Lukas2024SA}. The polariton dispersion directly encodes cyclotron dynamics, enabling extraction of the magnetic-field dependence of the cyclotron frequency; in particular, a square-root scaling with field would constitute a clear spectroscopic signature of linear (Dirac- or Weyl-like) band dispersions \cite{PbSnTe_CR,CR_sqrt_root}. The framework further allows coupling between cyclotron hyperbolic polaritons and other collective excitations, such as phonons and magnons, enriching the polaritonic spectrum and enabling these collective modes themselves to be accessed through polariton measurements. When combined with device engineering, such as lateral confinement to form polaritonic nanocavities \cite{HerzigSheinfux2024HBNcavities}, this platform naturally connects to cavity-QED in the terahertz regime and offers a versatile route to reconfigurable cavities and to exploring strong- and ultrastrong-coupling light–matter interactions.

In summary, we demonstrate that magnetic-field-driven hyperbolic polaritons can be realized in high-mobility semimetals. We establish a comprehensive theoretical framework for their formation and dispersion and propose experimentally accessible schemes for their detection and characterization. Our results position high-mobility semimetals as a versatile platform for magnetically tunable polaritonics, opening new opportunities for controlling light–matter interactions at the nanoscale.

\vspace{\baselineskip}

\textit{Acknowledgement} --- We would like to thank Fuyang Tay, Bo Zhao, Sina Jafari Ghalekohneh, Fl\'avio Feres, Susanne Stemmer for helpful discussion. We also thank N. P. Armitage for generously sharing magneto-terahertz conductivity data of Cd$_3$As$_2$ films. M.K.L. and B.C. acknowledge the Gordon and Betty Moore Foundation Grant DOI: 10.37807/gbmf12258 for supporting the development of polaritonic materials. M.K.L. acknowledges support for developing sensitive probes for electrodynamics in chiral materials from the NSF Faculty Early Career Development Program under Grant No. DMR—2045425.

\vspace{\baselineskip}

B.C. designed research; B.C. and Z.J.Z performed research and wrote the paper with input from all co-authors.

 \bibliography{Quadratic}

\begin{thebibliography}{40}
\expandafter\ifx\csname natexlab\endcsname\relax\def\natexlab#1{#1}\fi
\expandafter\ifx\csname bibnamefont\endcsname\relax
  \def\bibnamefont#1{#1}\fi
\expandafter\ifx\csname bibfnamefont\endcsname\relax
  \def\bibfnamefont#1{#1}\fi
\expandafter\ifx\csname citenamefont\endcsname\relax
  \def\citenamefont#1{#1}\fi
\expandafter\ifx\csname url\endcsname\relax
  \def\url#1{\texttt{#1}}\fi
\expandafter\ifx\csname urlprefix\endcsname\relax\def\urlprefix{URL }\fi
\providecommand{\bibinfo}[2]{#2}
\providecommand{\eprint}[2][]{\url{#2}}

\bibitem[{\citenamefont{Basov et~al.}(2016)\citenamefont{Basov, Fogler, and
  de~Abajo}}]{HP_Basov_review}
\bibinfo{author}{\bibfnamefont{D.~N.} \bibnamefont{Basov}},
  \bibinfo{author}{\bibfnamefont{M.~M.} \bibnamefont{Fogler}},
  \bibnamefont{and} \bibinfo{author}{\bibfnamefont{F.~J.~G.}
  \bibnamefont{de~Abajo}}, \bibinfo{journal}{Science}
  \textbf{\bibinfo{volume}{354}}, \bibinfo{pages}{aag1992}
  (\bibinfo{year}{2016}).

\bibitem[{\citenamefont{Low et~al.}(2017)\citenamefont{Low, Chaves, Caldwell,
  Kumar, Fang, Avouris, Heinz, Guinea, Martin-Moreno, and
  Koppens}}]{HP_Koppens_review}
\bibinfo{author}{\bibfnamefont{T.}~\bibnamefont{Low}},
  \bibinfo{author}{\bibfnamefont{A.}~\bibnamefont{Chaves}},
  \bibinfo{author}{\bibfnamefont{J.~D.} \bibnamefont{Caldwell}},
  \bibinfo{author}{\bibfnamefont{A.}~\bibnamefont{Kumar}},
  \bibinfo{author}{\bibfnamefont{N.~X.} \bibnamefont{Fang}},
  \bibinfo{author}{\bibfnamefont{P.}~\bibnamefont{Avouris}},
  \bibinfo{author}{\bibfnamefont{T.~F.} \bibnamefont{Heinz}},
  \bibinfo{author}{\bibfnamefont{F.}~\bibnamefont{Guinea}},
  \bibinfo{author}{\bibfnamefont{L.}~\bibnamefont{Martin-Moreno}},
  \bibnamefont{and} \bibinfo{author}{\bibfnamefont{F.}~\bibnamefont{Koppens}},
  \bibinfo{journal}{Nature Materials} \textbf{\bibinfo{volume}{16}},
  \bibinfo{pages}{182} (\bibinfo{year}{2017}).

\bibitem[{\citenamefont{Caldwell et~al.}(2014)\citenamefont{Caldwell, Kretinin,
  Chen, Giannini, Fogler, Francescato, Ellis, Tischler, Woods, Giles
  et~al.}}]{Caldwell2014}
\bibinfo{author}{\bibfnamefont{J.~D.} \bibnamefont{Caldwell}},
  \bibinfo{author}{\bibfnamefont{A.~V.} \bibnamefont{Kretinin}},
  \bibinfo{author}{\bibfnamefont{Y.}~\bibnamefont{Chen}},
  \bibinfo{author}{\bibfnamefont{V.}~\bibnamefont{Giannini}},
  \bibinfo{author}{\bibfnamefont{M.~M.} \bibnamefont{Fogler}},
  \bibinfo{author}{\bibfnamefont{Y.}~\bibnamefont{Francescato}},
  \bibinfo{author}{\bibfnamefont{C.~T.} \bibnamefont{Ellis}},
  \bibinfo{author}{\bibfnamefont{J.~G.} \bibnamefont{Tischler}},
  \bibinfo{author}{\bibfnamefont{C.~R.} \bibnamefont{Woods}},
  \bibinfo{author}{\bibfnamefont{A.~J.} \bibnamefont{Giles}},
  \bibnamefont{et~al.}, \bibinfo{journal}{Nature Communications}
  \textbf{\bibinfo{volume}{5}}, \bibinfo{pages}{5221} (\bibinfo{year}{2014}).

\bibitem[{\citenamefont{Caldwell et~al.}(2019)\citenamefont{Caldwell,
  Aharonovich, Cassabois, Edgar, Gil, and Basov}}]{Caldwell2019_hBN_review}
\bibinfo{author}{\bibfnamefont{J.~D.} \bibnamefont{Caldwell}},
  \bibinfo{author}{\bibfnamefont{I.}~\bibnamefont{Aharonovich}},
  \bibinfo{author}{\bibfnamefont{G.}~\bibnamefont{Cassabois}},
  \bibinfo{author}{\bibfnamefont{J.~H.} \bibnamefont{Edgar}},
  \bibinfo{author}{\bibfnamefont{B.}~\bibnamefont{Gil}}, \bibnamefont{and}
  \bibinfo{author}{\bibfnamefont{D.~N.} \bibnamefont{Basov}},
  \bibinfo{journal}{Nature Reviews Materials} \textbf{\bibinfo{volume}{4}},
  \bibinfo{pages}{552} (\bibinfo{year}{2019}).

\bibitem[{\citenamefont{Garc{\'\i}a~de Abajo
  et~al.}(2025)\citenamefont{Garc{\'\i}a~de Abajo, Basov, Koppens, Orsini,
  Ceccanti, Castilla, Cavicchi, Polini, Gon{\c{c}}alves, Costa
  et~al.}}]{Roadmap2DPhotonics2025}
\bibinfo{author}{\bibfnamefont{F.~J.} \bibnamefont{Garc{\'\i}a~de Abajo}},
  \bibinfo{author}{\bibfnamefont{D.~N.} \bibnamefont{Basov}},
  \bibinfo{author}{\bibfnamefont{F.~H.~L.} \bibnamefont{Koppens}},
  \bibinfo{author}{\bibfnamefont{L.}~\bibnamefont{Orsini}},
  \bibinfo{author}{\bibfnamefont{M.}~\bibnamefont{Ceccanti}},
  \bibinfo{author}{\bibfnamefont{S.}~\bibnamefont{Castilla}},
  \bibinfo{author}{\bibfnamefont{L.}~\bibnamefont{Cavicchi}},
  \bibinfo{author}{\bibfnamefont{M.}~\bibnamefont{Polini}},
  \bibinfo{author}{\bibfnamefont{P.~A.~D.} \bibnamefont{Gon{\c{c}}alves}},
  \bibinfo{author}{\bibfnamefont{A.~T.} \bibnamefont{Costa}},
  \bibnamefont{et~al.}, \bibinfo{journal}{ACS Photonics}
  \textbf{\bibinfo{volume}{12}}, \bibinfo{pages}{3961} (\bibinfo{year}{2025}).

\bibitem[{\citenamefont{Basov et~al.}(2025)\citenamefont{Basov, Asenjo-Garcia,
  Schuck, Zhu, Rubio, Cavalleri, Delor, Fogler, and Liu}}]{Basov_review_2025}
\bibinfo{author}{\bibfnamefont{D.}~\bibnamefont{Basov}},
  \bibinfo{author}{\bibfnamefont{A.}~\bibnamefont{Asenjo-Garcia}},
  \bibinfo{author}{\bibfnamefont{P.~J.} \bibnamefont{Schuck}},
  \bibinfo{author}{\bibfnamefont{X.}~\bibnamefont{Zhu}},
  \bibinfo{author}{\bibfnamefont{A.}~\bibnamefont{Rubio}},
  \bibinfo{author}{\bibfnamefont{A.}~\bibnamefont{Cavalleri}},
  \bibinfo{author}{\bibfnamefont{M.}~\bibnamefont{Delor}},
  \bibinfo{author}{\bibfnamefont{M.~M.} \bibnamefont{Fogler}},
  \bibnamefont{and} \bibinfo{author}{\bibfnamefont{M.}~\bibnamefont{Liu}},
  \bibinfo{journal}{Nanophotonics} \textbf{\bibinfo{volume}{14}},
  \bibinfo{pages}{3723} (\bibinfo{year}{2025}).

\bibitem[{\citenamefont{Ashida et~al.}(2023)\citenamefont{Ashida, \ifmmode
  \dot{I}\else \.{I}\fi{}mamo\ifmmode~\breve{g}\else \u{g}\fi{}lu, and
  Demler}}]{CavityQED_2023}
\bibinfo{author}{\bibfnamefont{Y.}~\bibnamefont{Ashida}},
  \bibinfo{author}{\bibfnamefont{A.~m.~c.} \bibnamefont{\ifmmode \dot{I}\else
  \.{I}\fi{}mamo\ifmmode~\breve{g}\else \u{g}\fi{}lu}}, \bibnamefont{and}
  \bibinfo{author}{\bibfnamefont{E.}~\bibnamefont{Demler}},
  \bibinfo{journal}{Phys. Rev. Lett.} \textbf{\bibinfo{volume}{130}},
  \bibinfo{pages}{216901} (\bibinfo{year}{2023}).

\bibitem[{\citenamefont{Riolo et~al.}(2025)\citenamefont{Riolo, Tomadin, Mazza,
  Asgari, MacDonald, and Polini}}]{Cavity_2025}
\bibinfo{author}{\bibfnamefont{R.}~\bibnamefont{Riolo}},
  \bibinfo{author}{\bibfnamefont{A.}~\bibnamefont{Tomadin}},
  \bibinfo{author}{\bibfnamefont{G.}~\bibnamefont{Mazza}},
  \bibinfo{author}{\bibfnamefont{R.}~\bibnamefont{Asgari}},
  \bibinfo{author}{\bibfnamefont{A.~H.} \bibnamefont{MacDonald}},
  \bibnamefont{and} \bibinfo{author}{\bibfnamefont{M.}~\bibnamefont{Polini}},
  \bibinfo{journal}{Proceedings of the National Academy of Sciences}
  \textbf{\bibinfo{volume}{122}}, \bibinfo{pages}{e2407995122}
  (\bibinfo{year}{2025}).

\bibitem[{\citenamefont{Kipp et~al.}(2025)\citenamefont{Kipp, Bretscher,
  Schulte, Herrmann, Kusyak, Day, Kesavan, Matsuyama, Li, Langner
  et~al.}}]{Kipp2025CavityVdW}
\bibinfo{author}{\bibfnamefont{G.}~\bibnamefont{Kipp}},
  \bibinfo{author}{\bibfnamefont{H.~M.} \bibnamefont{Bretscher}},
  \bibinfo{author}{\bibfnamefont{B.}~\bibnamefont{Schulte}},
  \bibinfo{author}{\bibfnamefont{D.}~\bibnamefont{Herrmann}},
  \bibinfo{author}{\bibfnamefont{K.}~\bibnamefont{Kusyak}},
  \bibinfo{author}{\bibfnamefont{M.~W.} \bibnamefont{Day}},
  \bibinfo{author}{\bibfnamefont{S.}~\bibnamefont{Kesavan}},
  \bibinfo{author}{\bibfnamefont{T.}~\bibnamefont{Matsuyama}},
  \bibinfo{author}{\bibfnamefont{X.}~\bibnamefont{Li}},
  \bibinfo{author}{\bibfnamefont{S.~M.} \bibnamefont{Langner}},
  \bibnamefont{et~al.}, \bibinfo{journal}{Nature Physics}
  \textbf{\bibinfo{volume}{21}}, \bibinfo{pages}{1926} (\bibinfo{year}{2025}).

\bibitem[{\citenamefont{Keren et~al.}(2025)\citenamefont{Keren, Webb, Zhang,
  Xu, Sun, Kim, Shin, Zhang, Zhang, Pereira et~al.}}]{Keren2025CavityAlteredSC}
\bibinfo{author}{\bibfnamefont{I.}~\bibnamefont{Keren}},
  \bibinfo{author}{\bibfnamefont{T.~A.} \bibnamefont{Webb}},
  \bibinfo{author}{\bibfnamefont{S.}~\bibnamefont{Zhang}},
  \bibinfo{author}{\bibfnamefont{J.}~\bibnamefont{Xu}},
  \bibinfo{author}{\bibfnamefont{D.}~\bibnamefont{Sun}},
  \bibinfo{author}{\bibfnamefont{B.~S.~Y.} \bibnamefont{Kim}},
  \bibinfo{author}{\bibfnamefont{D.}~\bibnamefont{Shin}},
  \bibinfo{author}{\bibfnamefont{S.~S.} \bibnamefont{Zhang}},
  \bibinfo{author}{\bibfnamefont{J.}~\bibnamefont{Zhang}},
  \bibinfo{author}{\bibfnamefont{G.}~\bibnamefont{Pereira}},
  \bibnamefont{et~al.}, \bibinfo{journal}{arXiv:2505.17378}
  (\bibinfo{year}{2025}).

\bibitem[{\citenamefont{Dai et~al.}(2014)\citenamefont{Dai, Fei, Ma, Rodin,
  Wagner, McLeod, Liu, Gannett, Regan, Watanabe et~al.}}]{hBN_HP_2014}
\bibinfo{author}{\bibfnamefont{S.}~\bibnamefont{Dai}},
  \bibinfo{author}{\bibfnamefont{Z.}~\bibnamefont{Fei}},
  \bibinfo{author}{\bibfnamefont{Q.}~\bibnamefont{Ma}},
  \bibinfo{author}{\bibfnamefont{A.~S.} \bibnamefont{Rodin}},
  \bibinfo{author}{\bibfnamefont{M.}~\bibnamefont{Wagner}},
  \bibinfo{author}{\bibfnamefont{A.~S.} \bibnamefont{McLeod}},
  \bibinfo{author}{\bibfnamefont{M.~K.} \bibnamefont{Liu}},
  \bibinfo{author}{\bibfnamefont{W.}~\bibnamefont{Gannett}},
  \bibinfo{author}{\bibfnamefont{W.}~\bibnamefont{Regan}},
  \bibinfo{author}{\bibfnamefont{K.}~\bibnamefont{Watanabe}},
  \bibnamefont{et~al.}, \bibinfo{journal}{Science}
  \textbf{\bibinfo{volume}{343}}, \bibinfo{pages}{1125} (\bibinfo{year}{2014}).

\bibitem[{\citenamefont{Ma et~al.}(2018)\citenamefont{Ma, Alonso-Gonz{\'a}lez,
  Li, Nikitin, Yuan, Mart{\'i}n-S{\'a}nchez, Taboada-Guti{\'e}rrez, Amenabar,
  Li, V{\'e}lez et~al.}}]{Ma2018_MoO3}
\bibinfo{author}{\bibfnamefont{W.}~\bibnamefont{Ma}},
  \bibinfo{author}{\bibfnamefont{P.}~\bibnamefont{Alonso-Gonz{\'a}lez}},
  \bibinfo{author}{\bibfnamefont{S.}~\bibnamefont{Li}},
  \bibinfo{author}{\bibfnamefont{A.~Y.} \bibnamefont{Nikitin}},
  \bibinfo{author}{\bibfnamefont{J.}~\bibnamefont{Yuan}},
  \bibinfo{author}{\bibfnamefont{J.}~\bibnamefont{Mart{\'i}n-S{\'a}nchez}},
  \bibinfo{author}{\bibfnamefont{J.}~\bibnamefont{Taboada-Guti{\'e}rrez}},
  \bibinfo{author}{\bibfnamefont{I.}~\bibnamefont{Amenabar}},
  \bibinfo{author}{\bibfnamefont{P.}~\bibnamefont{Li}},
  \bibinfo{author}{\bibfnamefont{S.}~\bibnamefont{V{\'e}lez}},
  \bibnamefont{et~al.}, \bibinfo{journal}{Nature}
  \textbf{\bibinfo{volume}{562}}, \bibinfo{pages}{557} (\bibinfo{year}{2018}).

\bibitem[{\citenamefont{Dai et~al.}(2015)\citenamefont{Dai, Ma, Liu, Andersen,
  Fei, Goldflam, Wagner, Watanabe, Taniguchi, Thiemens
  et~al.}}]{Dai2015_hBN_gra}
\bibinfo{author}{\bibfnamefont{S.}~\bibnamefont{Dai}},
  \bibinfo{author}{\bibfnamefont{Q.}~\bibnamefont{Ma}},
  \bibinfo{author}{\bibfnamefont{M.}~\bibnamefont{Liu}},
  \bibinfo{author}{\bibfnamefont{T.}~\bibnamefont{Andersen}},
  \bibinfo{author}{\bibfnamefont{Z.}~\bibnamefont{Fei}},
  \bibinfo{author}{\bibfnamefont{M.~D.} \bibnamefont{Goldflam}},
  \bibinfo{author}{\bibfnamefont{M.}~\bibnamefont{Wagner}},
  \bibinfo{author}{\bibfnamefont{K.}~\bibnamefont{Watanabe}},
  \bibinfo{author}{\bibfnamefont{T.}~\bibnamefont{Taniguchi}},
  \bibinfo{author}{\bibfnamefont{M.}~\bibnamefont{Thiemens}},
  \bibnamefont{et~al.}, \bibinfo{journal}{Nature Nanotechnology}
  \textbf{\bibinfo{volume}{10}}, \bibinfo{pages}{682} (\bibinfo{year}{2015}).

\bibitem[{\citenamefont{Woessner et~al.}(2015)\citenamefont{Woessner,
  Lundeberg, Gao, Principi, Alonso-Gonzalez, Carrega, Watanabe, Taniguchi,
  Vignale, Polini et~al.}}]{Woessner2015_hBN_gra}
\bibinfo{author}{\bibfnamefont{A.}~\bibnamefont{Woessner}},
  \bibinfo{author}{\bibfnamefont{M.~B.} \bibnamefont{Lundeberg}},
  \bibinfo{author}{\bibfnamefont{Y.}~\bibnamefont{Gao}},
  \bibinfo{author}{\bibfnamefont{A.}~\bibnamefont{Principi}},
  \bibinfo{author}{\bibfnamefont{P.}~\bibnamefont{Alonso-Gonzalez}},
  \bibinfo{author}{\bibfnamefont{M.}~\bibnamefont{Carrega}},
  \bibinfo{author}{\bibfnamefont{K.}~\bibnamefont{Watanabe}},
  \bibinfo{author}{\bibfnamefont{T.}~\bibnamefont{Taniguchi}},
  \bibinfo{author}{\bibfnamefont{G.}~\bibnamefont{Vignale}},
  \bibinfo{author}{\bibfnamefont{M.}~\bibnamefont{Polini}},
  \bibnamefont{et~al.}, \bibinfo{journal}{Nature Materials}
  \textbf{\bibinfo{volume}{14}}, \bibinfo{pages}{421} (\bibinfo{year}{2015}).

\bibitem[{\citenamefont{Hu et~al.}(2022)\citenamefont{Hu, Chen, Teng, Yu, Qu,
  Sun, Xue, Hu, Wu, Li et~al.}}]{Hu2022_Gra_MoO3}
\bibinfo{author}{\bibfnamefont{H.}~\bibnamefont{Hu}},
  \bibinfo{author}{\bibfnamefont{N.}~\bibnamefont{Chen}},
  \bibinfo{author}{\bibfnamefont{H.}~\bibnamefont{Teng}},
  \bibinfo{author}{\bibfnamefont{R.}~\bibnamefont{Yu}},
  \bibinfo{author}{\bibfnamefont{Y.}~\bibnamefont{Qu}},
  \bibinfo{author}{\bibfnamefont{J.}~\bibnamefont{Sun}},
  \bibinfo{author}{\bibfnamefont{M.}~\bibnamefont{Xue}},
  \bibinfo{author}{\bibfnamefont{D.}~\bibnamefont{Hu}},
  \bibinfo{author}{\bibfnamefont{B.}~\bibnamefont{Wu}},
  \bibinfo{author}{\bibfnamefont{C.}~\bibnamefont{Li}}, \bibnamefont{et~al.},
  \bibinfo{journal}{Nature Nanotechnology} \textbf{\bibinfo{volume}{17}},
  \bibinfo{pages}{940} (\bibinfo{year}{2022}).

\bibitem[{\citenamefont{Ruta et~al.}(2022)\citenamefont{Ruta, Kim, Sun, Rizzo,
  McLeod, Rajendran, Liu, Millis, Hone, and Basov}}]{Ruta2022}
\bibinfo{author}{\bibfnamefont{F.~L.} \bibnamefont{Ruta}},
  \bibinfo{author}{\bibfnamefont{B.~S.~Y.} \bibnamefont{Kim}},
  \bibinfo{author}{\bibfnamefont{Z.}~\bibnamefont{Sun}},
  \bibinfo{author}{\bibfnamefont{D.~J.} \bibnamefont{Rizzo}},
  \bibinfo{author}{\bibfnamefont{A.~S.} \bibnamefont{McLeod}},
  \bibinfo{author}{\bibfnamefont{A.}~\bibnamefont{Rajendran}},
  \bibinfo{author}{\bibfnamefont{S.}~\bibnamefont{Liu}},
  \bibinfo{author}{\bibfnamefont{A.~J.} \bibnamefont{Millis}},
  \bibinfo{author}{\bibfnamefont{J.~C.} \bibnamefont{Hone}}, \bibnamefont{and}
  \bibinfo{author}{\bibfnamefont{D.~N.} \bibnamefont{Basov}},
  \bibinfo{journal}{Nature Communications} \textbf{\bibinfo{volume}{13}},
  \bibinfo{pages}{3719} (\bibinfo{year}{2022}).

\bibitem[{\citenamefont{Sternbach et~al.}(2023)\citenamefont{Sternbach, Moore,
  Rikhter, Zhang, Jing, Shao, Kim, Xu, Liu, Edgar et~al.}}]{Sternbach2023}
\bibinfo{author}{\bibfnamefont{A.~J.} \bibnamefont{Sternbach}},
  \bibinfo{author}{\bibfnamefont{S.~L.} \bibnamefont{Moore}},
  \bibinfo{author}{\bibfnamefont{A.}~\bibnamefont{Rikhter}},
  \bibinfo{author}{\bibfnamefont{S.}~\bibnamefont{Zhang}},
  \bibinfo{author}{\bibfnamefont{R.}~\bibnamefont{Jing}},
  \bibinfo{author}{\bibfnamefont{Y.}~\bibnamefont{Shao}},
  \bibinfo{author}{\bibfnamefont{B.~S.~Y.} \bibnamefont{Kim}},
  \bibinfo{author}{\bibfnamefont{S.}~\bibnamefont{Xu}},
  \bibinfo{author}{\bibfnamefont{S.}~\bibnamefont{Liu}},
  \bibinfo{author}{\bibfnamefont{J.~H.} \bibnamefont{Edgar}},
  \bibnamefont{et~al.}, \bibinfo{journal}{Science}
  \textbf{\bibinfo{volume}{379}}, \bibinfo{pages}{555} (\bibinfo{year}{2023}).

\bibitem[{\citenamefont{Wehmeier
  et~al.}(2024{\natexlab{a}})\citenamefont{Wehmeier, Yu, Chen, Mayer, Xiong,
  Yao, Jiang, Hu, Janzen, Edgar et~al.}}]{Lukas2024}
\bibinfo{author}{\bibfnamefont{L.}~\bibnamefont{Wehmeier}},
  \bibinfo{author}{\bibfnamefont{S.-J.} \bibnamefont{Yu}},
  \bibinfo{author}{\bibfnamefont{X.}~\bibnamefont{Chen}},
  \bibinfo{author}{\bibfnamefont{R.~A.} \bibnamefont{Mayer}},
  \bibinfo{author}{\bibfnamefont{L.}~\bibnamefont{Xiong}},
  \bibinfo{author}{\bibfnamefont{H.}~\bibnamefont{Yao}},
  \bibinfo{author}{\bibfnamefont{Y.}~\bibnamefont{Jiang}},
  \bibinfo{author}{\bibfnamefont{J.}~\bibnamefont{Hu}},
  \bibinfo{author}{\bibfnamefont{E.}~\bibnamefont{Janzen}},
  \bibinfo{author}{\bibfnamefont{J.~H.} \bibnamefont{Edgar}},
  \bibnamefont{et~al.}, \bibinfo{journal}{Advanced Materials}
  \textbf{\bibinfo{volume}{36}}, \bibinfo{pages}{2401349}
  (\bibinfo{year}{2024}{\natexlab{a}}).

\bibitem[{\citenamefont{Shao et~al.}(2022)\citenamefont{Shao, Sternbach, Kim,
  Rikhter, Xu, Giovannini, Jing, Chae, Sun, Lee et~al.}}]{HP_nodal_metal_2022}
\bibinfo{author}{\bibfnamefont{Y.}~\bibnamefont{Shao}},
  \bibinfo{author}{\bibfnamefont{A.~J.} \bibnamefont{Sternbach}},
  \bibinfo{author}{\bibfnamefont{B.~S.~Y.} \bibnamefont{Kim}},
  \bibinfo{author}{\bibfnamefont{A.~A.} \bibnamefont{Rikhter}},
  \bibinfo{author}{\bibfnamefont{X.}~\bibnamefont{Xu}},
  \bibinfo{author}{\bibfnamefont{U.~D.} \bibnamefont{Giovannini}},
  \bibinfo{author}{\bibfnamefont{R.}~\bibnamefont{Jing}},
  \bibinfo{author}{\bibfnamefont{S.~H.} \bibnamefont{Chae}},
  \bibinfo{author}{\bibfnamefont{Z.}~\bibnamefont{Sun}},
  \bibinfo{author}{\bibfnamefont{S.~H.} \bibnamefont{Lee}},
  \bibnamefont{et~al.}, \bibinfo{journal}{Science Advances}
  \textbf{\bibinfo{volume}{8}}, \bibinfo{pages}{eadd6169}
  (\bibinfo{year}{2022}).

\bibitem[{\citenamefont{Ruta et~al.}(2025)\citenamefont{Ruta, Shao, Acharya,
  Mu, Jo, Ryu, Balatsky, Su, Pashov, Kim et~al.}}]{HP_metal_2025}
\bibinfo{author}{\bibfnamefont{F.~L.} \bibnamefont{Ruta}},
  \bibinfo{author}{\bibfnamefont{Y.}~\bibnamefont{Shao}},
  \bibinfo{author}{\bibfnamefont{S.}~\bibnamefont{Acharya}},
  \bibinfo{author}{\bibfnamefont{A.}~\bibnamefont{Mu}},
  \bibinfo{author}{\bibfnamefont{N.~H.} \bibnamefont{Jo}},
  \bibinfo{author}{\bibfnamefont{S.~H.} \bibnamefont{Ryu}},
  \bibinfo{author}{\bibfnamefont{D.}~\bibnamefont{Balatsky}},
  \bibinfo{author}{\bibfnamefont{Y.}~\bibnamefont{Su}},
  \bibinfo{author}{\bibfnamefont{D.}~\bibnamefont{Pashov}},
  \bibinfo{author}{\bibfnamefont{B.~S.~Y.} \bibnamefont{Kim}},
  \bibnamefont{et~al.}, \bibinfo{journal}{Science}
  \textbf{\bibinfo{volume}{387}}, \bibinfo{pages}{786} (\bibinfo{year}{2025}).

\bibitem[{\citenamefont{Venturi et~al.}(2024)\citenamefont{Venturi, Mancini,
  Melchioni, Chiodini, and Ambrosio}}]{Venturi2024}
\bibinfo{author}{\bibfnamefont{G.}~\bibnamefont{Venturi}},
  \bibinfo{author}{\bibfnamefont{A.}~\bibnamefont{Mancini}},
  \bibinfo{author}{\bibfnamefont{N.}~\bibnamefont{Melchioni}},
  \bibinfo{author}{\bibfnamefont{S.}~\bibnamefont{Chiodini}}, \bibnamefont{and}
  \bibinfo{author}{\bibfnamefont{A.}~\bibnamefont{Ambrosio}},
  \bibinfo{journal}{Nature Communications} \textbf{\bibinfo{volume}{15}},
  \bibinfo{pages}{9727} (\bibinfo{year}{2024}).

\bibitem[{\citenamefont{Wu et~al.}(2015)\citenamefont{Wu, Tse, Brahlek, Morris,
  Aguilar, Koirala, Oh, and Armitage}}]{Bi2Se3_CR}
\bibinfo{author}{\bibfnamefont{L.}~\bibnamefont{Wu}},
  \bibinfo{author}{\bibfnamefont{W.-K.} \bibnamefont{Tse}},
  \bibinfo{author}{\bibfnamefont{M.}~\bibnamefont{Brahlek}},
  \bibinfo{author}{\bibfnamefont{C.~M.} \bibnamefont{Morris}},
  \bibinfo{author}{\bibfnamefont{R.~V.} \bibnamefont{Aguilar}},
  \bibinfo{author}{\bibfnamefont{N.}~\bibnamefont{Koirala}},
  \bibinfo{author}{\bibfnamefont{S.}~\bibnamefont{Oh}}, \bibnamefont{and}
  \bibinfo{author}{\bibfnamefont{N.~P.} \bibnamefont{Armitage}},
  \bibinfo{journal}{Phys. Rev. Lett.} \textbf{\bibinfo{volume}{115}},
  \bibinfo{pages}{217602} (\bibinfo{year}{2015}).

\bibitem[{\citenamefont{Akrap et~al.}(2016)\citenamefont{Akrap, Hakl,
  Tchoumakov, Crassee, Kuba, Goerbig, Homes, Caha, Nov\'ak, Teppe
  et~al.}}]{CdAs_IR_magneto2016}
\bibinfo{author}{\bibfnamefont{A.}~\bibnamefont{Akrap}},
  \bibinfo{author}{\bibfnamefont{M.}~\bibnamefont{Hakl}},
  \bibinfo{author}{\bibfnamefont{S.}~\bibnamefont{Tchoumakov}},
  \bibinfo{author}{\bibfnamefont{I.}~\bibnamefont{Crassee}},
  \bibinfo{author}{\bibfnamefont{J.}~\bibnamefont{Kuba}},
  \bibinfo{author}{\bibfnamefont{M.~O.} \bibnamefont{Goerbig}},
  \bibinfo{author}{\bibfnamefont{C.~C.} \bibnamefont{Homes}},
  \bibinfo{author}{\bibfnamefont{O.}~\bibnamefont{Caha}},
  \bibinfo{author}{\bibfnamefont{J.}~\bibnamefont{Nov\'ak}},
  \bibinfo{author}{\bibfnamefont{F.}~\bibnamefont{Teppe}},
  \bibnamefont{et~al.}, \bibinfo{journal}{Phys. Rev. Lett.}
  \textbf{\bibinfo{volume}{117}}, \bibinfo{pages}{136401}
  (\bibinfo{year}{2016}).

\bibitem[{\citenamefont{Cheng et~al.}(2019)\citenamefont{Cheng, Taylor, Folkes,
  Rong, and Armitage}}]{PbSnTe_CR}
\bibinfo{author}{\bibfnamefont{B.}~\bibnamefont{Cheng}},
  \bibinfo{author}{\bibfnamefont{P.}~\bibnamefont{Taylor}},
  \bibinfo{author}{\bibfnamefont{P.}~\bibnamefont{Folkes}},
  \bibinfo{author}{\bibfnamefont{C.}~\bibnamefont{Rong}}, \bibnamefont{and}
  \bibinfo{author}{\bibfnamefont{N.~P.} \bibnamefont{Armitage}},
  \bibinfo{journal}{Phys. Rev. Lett.} \textbf{\bibinfo{volume}{122}},
  \bibinfo{pages}{097401} (\bibinfo{year}{2019}).

\bibitem[{\citenamefont{Cheng et~al.}(2020)\citenamefont{Cheng, Schumann, Wang,
  Zhang, Barbalas, Stemmer, and Armitage}}]{CdAs_CR}
\bibinfo{author}{\bibfnamefont{B.}~\bibnamefont{Cheng}},
  \bibinfo{author}{\bibfnamefont{T.}~\bibnamefont{Schumann}},
  \bibinfo{author}{\bibfnamefont{Y.}~\bibnamefont{Wang}},
  \bibinfo{author}{\bibfnamefont{X.}~\bibnamefont{Zhang}},
  \bibinfo{author}{\bibfnamefont{D.}~\bibnamefont{Barbalas}},
  \bibinfo{author}{\bibfnamefont{S.}~\bibnamefont{Stemmer}}, \bibnamefont{and}
  \bibinfo{author}{\bibfnamefont{N.~P.} \bibnamefont{Armitage}},
  \bibinfo{journal}{Nano Letters} \textbf{\bibinfo{volume}{20}},
  \bibinfo{pages}{5991} (\bibinfo{year}{2020}).

\bibitem[{\citenamefont{Xiong et~al.}(2015)\citenamefont{Xiong, Kushwaha,
  Liang, Krizan, Hirschberger, Wang, Cava, and Ong}}]{Ong2015}
\bibinfo{author}{\bibfnamefont{J.}~\bibnamefont{Xiong}},
  \bibinfo{author}{\bibfnamefont{S.~K.} \bibnamefont{Kushwaha}},
  \bibinfo{author}{\bibfnamefont{T.}~\bibnamefont{Liang}},
  \bibinfo{author}{\bibfnamefont{J.~W.} \bibnamefont{Krizan}},
  \bibinfo{author}{\bibfnamefont{M.}~\bibnamefont{Hirschberger}},
  \bibinfo{author}{\bibfnamefont{W.}~\bibnamefont{Wang}},
  \bibinfo{author}{\bibfnamefont{R.~J.} \bibnamefont{Cava}}, \bibnamefont{and}
  \bibinfo{author}{\bibfnamefont{N.~P.} \bibnamefont{Ong}},
  \bibinfo{journal}{Science} \textbf{\bibinfo{volume}{350}},
  \bibinfo{pages}{413} (\bibinfo{year}{2015}).

\bibitem[{\citenamefont{Li et~al.}(2015)\citenamefont{Li, Wang, Liu, Wang,
  Liao, and Yu}}]{Li02015}
\bibinfo{author}{\bibfnamefont{C.-Z.} \bibnamefont{Li}},
  \bibinfo{author}{\bibfnamefont{L.-X.} \bibnamefont{Wang}},
  \bibinfo{author}{\bibfnamefont{H.}~\bibnamefont{Liu}},
  \bibinfo{author}{\bibfnamefont{J.}~\bibnamefont{Wang}},
  \bibinfo{author}{\bibfnamefont{Z.-M.} \bibnamefont{Liao}}, \bibnamefont{and}
  \bibinfo{author}{\bibfnamefont{D.-P.} \bibnamefont{Yu}},
  \bibinfo{journal}{Nature Communications} \textbf{\bibinfo{volume}{6}},
  \bibinfo{pages}{10137} (\bibinfo{year}{2015}).

\bibitem[{\citenamefont{Cheng et~al.}(2021)\citenamefont{Cheng, Schumann,
  Stemmer, and Armitage}}]{Cheng_2021}
\bibinfo{author}{\bibfnamefont{B.}~\bibnamefont{Cheng}},
  \bibinfo{author}{\bibfnamefont{T.}~\bibnamefont{Schumann}},
  \bibinfo{author}{\bibfnamefont{S.}~\bibnamefont{Stemmer}}, \bibnamefont{and}
  \bibinfo{author}{\bibfnamefont{N.~P.} \bibnamefont{Armitage}},
  \bibinfo{journal}{Science Advances} \textbf{\bibinfo{volume}{7}},
  \bibinfo{pages}{eabg0914} (\bibinfo{year}{2021}).

\bibitem[{\citenamefont{Gon{\c{c}}alves and Peres}(2016)}]{GoncalvesPeres2016}
\bibinfo{author}{\bibfnamefont{P.~A.~D.} \bibnamefont{Gon{\c{c}}alves}}
  \bibnamefont{and} \bibinfo{author}{\bibfnamefont{N.~M.~R.}
  \bibnamefont{Peres}}, \emph{\bibinfo{title}{An Introduction to Graphene
  Plasmonics}} (\bibinfo{publisher}{World Scientific},
  \bibinfo{address}{Singapore}, \bibinfo{year}{2016}).

\bibitem[{\citenamefont{Homes et~al.}(2015)\citenamefont{Homes, Ali, and
  Cava}}]{WTe2_IR_phonon_2015}
\bibinfo{author}{\bibfnamefont{C.~C.} \bibnamefont{Homes}},
  \bibinfo{author}{\bibfnamefont{M.~N.} \bibnamefont{Ali}}, \bibnamefont{and}
  \bibinfo{author}{\bibfnamefont{R.~J.} \bibnamefont{Cava}},
  \bibinfo{journal}{Phys. Rev. B} \textbf{\bibinfo{volume}{92}},
  \bibinfo{pages}{161109} (\bibinfo{year}{2015}).

\bibitem[{\citenamefont{Xu et~al.}(2018)\citenamefont{Xu, Zhao, Marsik,
  Sheveleva, Lyzwa, Dai, Chen, Qiu, and Bernhard}}]{ZrTe5_phonon_2018}
\bibinfo{author}{\bibfnamefont{B.}~\bibnamefont{Xu}},
  \bibinfo{author}{\bibfnamefont{L.~X.} \bibnamefont{Zhao}},
  \bibinfo{author}{\bibfnamefont{P.}~\bibnamefont{Marsik}},
  \bibinfo{author}{\bibfnamefont{E.}~\bibnamefont{Sheveleva}},
  \bibinfo{author}{\bibfnamefont{F.}~\bibnamefont{Lyzwa}},
  \bibinfo{author}{\bibfnamefont{Y.~M.} \bibnamefont{Dai}},
  \bibinfo{author}{\bibfnamefont{G.~F.} \bibnamefont{Chen}},
  \bibinfo{author}{\bibfnamefont{X.~G.} \bibnamefont{Qiu}}, \bibnamefont{and}
  \bibinfo{author}{\bibfnamefont{C.}~\bibnamefont{Bernhard}},
  \bibinfo{journal}{Phys. Rev. Lett.} \textbf{\bibinfo{volume}{121}},
  \bibinfo{pages}{187401} (\bibinfo{year}{2018}).

\bibitem[{\citenamefont{Xu et~al.}(2024)\citenamefont{Xu, Li, Vitalone, Jing,
  Sternbach, Zhang, Ingham, Delor, McIver, Yankowitz
  et~al.}}]{Map_SPP_THz_2024}
\bibinfo{author}{\bibfnamefont{S.}~\bibnamefont{Xu}},
  \bibinfo{author}{\bibfnamefont{Y.}~\bibnamefont{Li}},
  \bibinfo{author}{\bibfnamefont{R.~A.} \bibnamefont{Vitalone}},
  \bibinfo{author}{\bibfnamefont{R.}~\bibnamefont{Jing}},
  \bibinfo{author}{\bibfnamefont{A.~J.} \bibnamefont{Sternbach}},
  \bibinfo{author}{\bibfnamefont{S.}~\bibnamefont{Zhang}},
  \bibinfo{author}{\bibfnamefont{J.}~\bibnamefont{Ingham}},
  \bibinfo{author}{\bibfnamefont{M.}~\bibnamefont{Delor}},
  \bibinfo{author}{\bibfnamefont{J.~W.} \bibnamefont{McIver}},
  \bibinfo{author}{\bibfnamefont{M.}~\bibnamefont{Yankowitz}},
  \bibnamefont{et~al.}, \bibinfo{journal}{Science Advances}
  \textbf{\bibinfo{volume}{10}}, \bibinfo{pages}{eado5553}
  (\bibinfo{year}{2024}).

\bibitem[{\citenamefont{Yoxall et~al.}(2015)\citenamefont{Yoxall, Schnell,
  Nikitin, Txoperena, Woessner, Lundeberg, Casanova, Hueso, Koppens, and
  Hillenbrand}}]{Map_HP_2015}
\bibinfo{author}{\bibfnamefont{E.}~\bibnamefont{Yoxall}},
  \bibinfo{author}{\bibfnamefont{M.}~\bibnamefont{Schnell}},
  \bibinfo{author}{\bibfnamefont{A.~Y.} \bibnamefont{Nikitin}},
  \bibinfo{author}{\bibfnamefont{O.}~\bibnamefont{Txoperena}},
  \bibinfo{author}{\bibfnamefont{A.}~\bibnamefont{Woessner}},
  \bibinfo{author}{\bibfnamefont{M.~B.} \bibnamefont{Lundeberg}},
  \bibinfo{author}{\bibfnamefont{F.}~\bibnamefont{Casanova}},
  \bibinfo{author}{\bibfnamefont{L.~E.} \bibnamefont{Hueso}},
  \bibinfo{author}{\bibfnamefont{F.~H.~L.} \bibnamefont{Koppens}},
  \bibnamefont{and}
  \bibinfo{author}{\bibfnamefont{R.}~\bibnamefont{Hillenbrand}},
  \bibinfo{journal}{Nature Photonics} \textbf{\bibinfo{volume}{9}},
  \bibinfo{pages}{674} (\bibinfo{year}{2015}).

\bibitem[{\citenamefont{Liu et~al.}(2014{\natexlab{a}})\citenamefont{Liu, Zhou,
  Zhang, Wang, Weng, Prabhakaran, Mo, Shen, Fang, Dai et~al.}}]{Cd3As22013}
\bibinfo{author}{\bibfnamefont{Z.~K.} \bibnamefont{Liu}},
  \bibinfo{author}{\bibfnamefont{B.}~\bibnamefont{Zhou}},
  \bibinfo{author}{\bibfnamefont{Y.}~\bibnamefont{Zhang}},
  \bibinfo{author}{\bibfnamefont{Z.~J.} \bibnamefont{Wang}},
  \bibinfo{author}{\bibfnamefont{H.~M.} \bibnamefont{Weng}},
  \bibinfo{author}{\bibfnamefont{D.}~\bibnamefont{Prabhakaran}},
  \bibinfo{author}{\bibfnamefont{S.-K.} \bibnamefont{Mo}},
  \bibinfo{author}{\bibfnamefont{Z.~X.} \bibnamefont{Shen}},
  \bibinfo{author}{\bibfnamefont{Z.}~\bibnamefont{Fang}},
  \bibinfo{author}{\bibfnamefont{X.}~\bibnamefont{Dai}}, \bibnamefont{et~al.},
  \bibinfo{journal}{Science} \textbf{\bibinfo{volume}{343}},
  \bibinfo{pages}{864} (\bibinfo{year}{2014}{\natexlab{a}}).

\bibitem[{\citenamefont{Liu et~al.}(2014{\natexlab{b}})\citenamefont{Liu,
  Jiang, Zhou, Wang, Zhang, Weng, Prabhakaran, Mo, Peng, Dudin
  et~al.}}]{Liu2014}
\bibinfo{author}{\bibfnamefont{Z.~K.} \bibnamefont{Liu}},
  \bibinfo{author}{\bibfnamefont{J.}~\bibnamefont{Jiang}},
  \bibinfo{author}{\bibfnamefont{B.}~\bibnamefont{Zhou}},
  \bibinfo{author}{\bibfnamefont{Z.~J.} \bibnamefont{Wang}},
  \bibinfo{author}{\bibfnamefont{Y.}~\bibnamefont{Zhang}},
  \bibinfo{author}{\bibfnamefont{H.~M.} \bibnamefont{Weng}},
  \bibinfo{author}{\bibfnamefont{D.}~\bibnamefont{Prabhakaran}},
  \bibinfo{author}{\bibfnamefont{S.-K.} \bibnamefont{Mo}},
  \bibinfo{author}{\bibfnamefont{H.}~\bibnamefont{Peng}},
  \bibinfo{author}{\bibfnamefont{P.}~\bibnamefont{Dudin}},
  \bibnamefont{et~al.}, \bibinfo{journal}{Nature Materials}
  \textbf{\bibinfo{volume}{13}}, \bibinfo{pages}{677}
  (\bibinfo{year}{2014}{\natexlab{b}}).

\bibitem[{\citenamefont{Xu et~al.}(2015)\citenamefont{Xu, Belopolski, Alidoust,
  Neupane, Bian, Zhang, Sankar, Chang, Yuan, Lee et~al.}}]{Suyang2015}
\bibinfo{author}{\bibfnamefont{S.-Y.} \bibnamefont{Xu}},
  \bibinfo{author}{\bibfnamefont{I.}~\bibnamefont{Belopolski}},
  \bibinfo{author}{\bibfnamefont{N.}~\bibnamefont{Alidoust}},
  \bibinfo{author}{\bibfnamefont{M.}~\bibnamefont{Neupane}},
  \bibinfo{author}{\bibfnamefont{G.}~\bibnamefont{Bian}},
  \bibinfo{author}{\bibfnamefont{C.}~\bibnamefont{Zhang}},
  \bibinfo{author}{\bibfnamefont{R.}~\bibnamefont{Sankar}},
  \bibinfo{author}{\bibfnamefont{G.}~\bibnamefont{Chang}},
  \bibinfo{author}{\bibfnamefont{Z.}~\bibnamefont{Yuan}},
  \bibinfo{author}{\bibfnamefont{C.-C.} \bibnamefont{Lee}},
  \bibnamefont{et~al.}, \bibinfo{journal}{Science}
  \textbf{\bibinfo{volume}{349}}, \bibinfo{pages}{613} (\bibinfo{year}{2015}).

\bibitem[{\citenamefont{Armitage et~al.}(2018)\citenamefont{Armitage, Mele, and
  Vishwanath}}]{topo_semimetal_review}
\bibinfo{author}{\bibfnamefont{N.~P.} \bibnamefont{Armitage}},
  \bibinfo{author}{\bibfnamefont{E.~J.} \bibnamefont{Mele}}, \bibnamefont{and}
  \bibinfo{author}{\bibfnamefont{A.}~\bibnamefont{Vishwanath}},
  \bibinfo{journal}{Rev. Mod. Phys.} \textbf{\bibinfo{volume}{90}},
  \bibinfo{pages}{015001} (\bibinfo{year}{2018}).

\bibitem[{\citenamefont{Wehmeier
  et~al.}(2024{\natexlab{b}})\citenamefont{Wehmeier, Xu, Mayer, Vermilyea,
  Tsuneto, Dapolito, Pu, Du, Chen, Zheng et~al.}}]{Lukas2024SA}
\bibinfo{author}{\bibfnamefont{L.}~\bibnamefont{Wehmeier}},
  \bibinfo{author}{\bibfnamefont{S.}~\bibnamefont{Xu}},
  \bibinfo{author}{\bibfnamefont{R.~A.} \bibnamefont{Mayer}},
  \bibinfo{author}{\bibfnamefont{B.}~\bibnamefont{Vermilyea}},
  \bibinfo{author}{\bibfnamefont{M.}~\bibnamefont{Tsuneto}},
  \bibinfo{author}{\bibfnamefont{M.}~\bibnamefont{Dapolito}},
  \bibinfo{author}{\bibfnamefont{R.}~\bibnamefont{Pu}},
  \bibinfo{author}{\bibfnamefont{Z.}~\bibnamefont{Du}},
  \bibinfo{author}{\bibfnamefont{X.}~\bibnamefont{Chen}},
  \bibinfo{author}{\bibfnamefont{W.}~\bibnamefont{Zheng}},
  \bibnamefont{et~al.}, \bibinfo{journal}{Science Advances}
  \textbf{\bibinfo{volume}{10}}, \bibinfo{pages}{eadp3487}
  (\bibinfo{year}{2024}{\natexlab{b}}).

\bibitem[{\citenamefont{Shao et~al.}(2024)\citenamefont{Shao, Moon, Rudenko,
  Wang, Herzog-Arbeitman, Ozerov, Graf, Sun, Queiroz, Lee
  et~al.}}]{CR_sqrt_root}
\bibinfo{author}{\bibfnamefont{Y.}~\bibnamefont{Shao}},
  \bibinfo{author}{\bibfnamefont{S.}~\bibnamefont{Moon}},
  \bibinfo{author}{\bibfnamefont{A.~N.} \bibnamefont{Rudenko}},
  \bibinfo{author}{\bibfnamefont{J.}~\bibnamefont{Wang}},
  \bibinfo{author}{\bibfnamefont{J.}~\bibnamefont{Herzog-Arbeitman}},
  \bibinfo{author}{\bibfnamefont{M.}~\bibnamefont{Ozerov}},
  \bibinfo{author}{\bibfnamefont{D.}~\bibnamefont{Graf}},
  \bibinfo{author}{\bibfnamefont{Z.}~\bibnamefont{Sun}},
  \bibinfo{author}{\bibfnamefont{R.}~\bibnamefont{Queiroz}},
  \bibinfo{author}{\bibfnamefont{S.~H.} \bibnamefont{Lee}},
  \bibnamefont{et~al.}, \bibinfo{journal}{Phys. Rev. X}
  \textbf{\bibinfo{volume}{14}}, \bibinfo{pages}{041057}
  (\bibinfo{year}{2024}).

\bibitem[{\citenamefont{Herzig~Sheinfux
  et~al.}(2024)\citenamefont{Herzig~Sheinfux, Orsini, Jung, Torre, Ceccanti,
  Marconi, Maniyara, Barcons~Ruiz, H{\"o}tger, Bertini
  et~al.}}]{HerzigSheinfux2024HBNcavities}
\bibinfo{author}{\bibfnamefont{H.}~\bibnamefont{Herzig~Sheinfux}},
  \bibinfo{author}{\bibfnamefont{L.}~\bibnamefont{Orsini}},
  \bibinfo{author}{\bibfnamefont{M.}~\bibnamefont{Jung}},
  \bibinfo{author}{\bibfnamefont{I.}~\bibnamefont{Torre}},
  \bibinfo{author}{\bibfnamefont{M.}~\bibnamefont{Ceccanti}},
  \bibinfo{author}{\bibfnamefont{S.}~\bibnamefont{Marconi}},
  \bibinfo{author}{\bibfnamefont{R.}~\bibnamefont{Maniyara}},
  \bibinfo{author}{\bibfnamefont{D.}~\bibnamefont{Barcons~Ruiz}},
  \bibinfo{author}{\bibfnamefont{A.}~\bibnamefont{H{\"o}tger}},
  \bibinfo{author}{\bibfnamefont{R.}~\bibnamefont{Bertini}},
  \bibnamefont{et~al.}, \bibinfo{journal}{Nature Materials}
  \textbf{\bibinfo{volume}{23}}, \bibinfo{pages}{499} (\bibinfo{year}{2024}).

\end{thebibliography}
 
\end{document}